\documentclass[aps,pre,preprint,groupedaddress,showpacs]{revtex4-1}

\usepackage{graphics}
\usepackage{latexsym}
\usepackage{bm}
\usepackage{fancybox}
\usepackage{accents}
\usepackage{mathrsfs}
\usepackage{mathrsfs}

\newcommand{\grad}{\nabla}

\newcommand{\wz}{\omega}

\newcommand{\hwz}{\hat{\omega}}

\newcommand{\dwz}{\delta\omega}
\renewcommand{\vr}{\vec{r}}
\newcommand{\vk}{\vec{k}}

\newcommand{\vq}{\vec{q}}
\renewcommand{\div}{\nabla \cdot}
\renewcommand{\u}{\vec{u}}

\renewcommand{\vec}[1]{\bm{#1}}

\begin{document}


\title{Explicit formula of energy-conserving, Fokker-Planck type
collision term for nonaxisymmetric, single species point vortex system}


\author{Yuichi Yatsuyanagi}
\affiliation{Faculty of Education, Shizuoka University, Suruga-ku, Shizuoka 422-8529, Japan}
\author{Tadatsugu Hatori}
\affiliation{National Institute for Fusion Science, Toki, Gifu 509-5292, Japan}


\date{\today}

\begin{abstract}
This paper considers a kinetic equation for an unbounded two-dimensional
single species point vortex system.
No axisymmetric flow is assumed.
Using the kinetic theory based on the Klimontovich formalism, we derive
a collision term consisting of a diffusion and a drift term, whose structure
is similar to the Fokker-Planck equation.
The collision term exhibits several important properties:
(a) it includes a nonlocal effect;
(b) it conserves the mean field energy; 
(c) it satisfies the H theorem;
(d) its effect vanishes in each local equilibrium region with the same temperature.
When the system reaches a global equilibrium state, the collision term completely
converge to zero all over the system.
\end{abstract}



\maketitle

\section{Introduction}
Two-dimensional (2D) microscopic point vortex is a formal solution
of the 2D inviscid microscopic Euler equation, 
\begin{equation}
	\frac{\partial }{\partial t} \hwz(\vr,t) + \div (\hat{\u}(\vr,t) \hwz(\vr,t)) = 0 \label{eqn:euler}
\end{equation}
where $\hwz(\vr,t)$ and $\hat{\u}(\vr,t)$ are the microscopic vorticity and the microscopic
velocity, respectively.
Equation (\ref{eqn:euler}) is formally identical with the macroscopic
Euler equation
\begin{equation}
	\frac{\partial }{\partial t} \wz(\vr,t) + \div (\u(\vr,t) \wz(\vr,t)) = 0 \label{eqn:macroeuler}
\end{equation}
where $\wz(\vr,t) \equiv \langle\hwz(\vr,t)\rangle$ and $\u(\vr,t)\equiv\langle\hat{\u}(\vr,t)\rangle$ 
are the macroscopic vorticity and the macroscopic velocity, respectively.
Operator $\langle \cdot \rangle$ is an averaging operator.
The point vortex system has been successfully applied
to the study of 2D turbulence \cite{Kraichnan, Tabeling}.
In the landmark paper published in 1949, Onsager proposed 
an application of statistical mechanics to 2D point 
vortex system, in which he sketched a possible explanation for 
the formation of large-scale, long-lived vortex structures in 
turbulent flows \cite{Onsager, Eyink}.
Equilibrium distribution at negative temperature described by the sinh-Poisson
equation is found by Joyce and Montgomery \cite{Joyce}.
Since then, large research effort has been devoted to understand the
negative temperature state, both theoretically and numerically
\cite{Montgomery1974, Kida1975, Kraichnan1975, Seyler1976, Pointin1976, Lundgren1977, Ting1987, Robert1991, Eyink1993, Buehler2002, Yatsuyanagi2005}.
On the other hand, it has been pointed out that a decaying, 2D 
Navier-Stokes turbulence reaches an equilibrium state described 
by sinh-Poisson equation \cite{Matthaeus1991, Montgomery1993, Li1996}.

Here, a question arises.
A distribution of the point vortices $\hwz(\vr,0)$ 
is given at $t=0$.
A time-evolved distribution $\hwz(\vr, T)$ 
at certain time $T$ 
is obtained by solving the microscopic Euler equation (\ref{eqn:euler}).
On the other hand, a macroscopic vorticity field $\omega(\vr,0)$ at $t=0$ 
is obtained by a space average of $\hwz(\vr, 0)$, 
namely $\omega(\vr,0) = \langle \hwz(\vr,0) \rangle$.
Of course, a time-evolved macroscopic vorticity field
$\omega(\vr, T)$ is obtained by solving the macroscopic Euler equation 
(\ref{eqn:macroeuler}).
Is the space-averaged point vortex solution $\langle \hwz(\vr,T) \rangle$
the same as the macroscopic vorticity filed $\omega(\vr,T)$?
We believe that the answer is ``no" and the evolving equation
for $\omega(\vr,T)$ which is exactly equal to $\langle \hwz(\vr,T) \rangle$ 
should be written as 
\begin{equation}
	\frac{\partial \wz(\vr,t)}{\partial t} + \div (\u(\vr,t) \wz(\vr,t)) = C \label{eqn:macro-with-collision}
\end{equation}
where $C$ is a collision term.
In the following, we restrict our discussion to determining 
explicit formula of $C$, i.e., a kinetic theory with the Klimontovich 
formalism \cite{Klimontovich}.

In the context of relaxation processes of the point vortex system
towards a statistical equilibrium state, diffusive effect 
due to the discrete distribution has attracted a lot of attention.
The Fokker-Planck type collision term consisting of the diffusion term 
and the drift (friction) term was discussed by Dubin and O'Neil, in which
an axially symmetric velocity profile is assumed and a collective effect 
is formally taken into account \cite{DubinONeil1988,Dubin2003}.
This result yields zero diffusion flux for single species point
vortices.
On the other hand, Chavanis obtained the kinetic equation
with a collective effect including the implicit term \cite{Chavanis2001,Chavanis2012}.
By dropping the collective effect, the explicit formula is obtained.
However, the result has a problem that a monotonic angular velocity profile 
yields zero collisional effect and the system does not reach a thermal 
equilibrium state.

In this paper, we present a kinetic equation for the unbounded, 
2D single species point vortex system.
To derive an explicit formula of the collision term, Klimontovich 
formalism is used.
The obtained collision term has following good properties that resolve 
the above issues.
The collision term conserves the mean field energy.
During a relaxation process towards the global equilibrium,
system reaches a local equilibrium state first.
In the local equilibrium state, the relation 
$\omega_{\rm leq} = \omega[\psi]$
is satisfied in small regions where $\omega[\psi]$ is a functional
of the stream function $\psi$ and in the regions, the second term
$\div (\u \omega)$ of the Euler equation  (\ref{eqn:macro-with-collision}) 
vanishes.
Then time evolution of the system is dominated by the collision
term.
However, the magnitude of the collision term is small compared
to that of $\div (\u\omega)$, and the speed of the relaxation
slows down.
When system reaches a global equilibrium state described by
the Boltzmann distribution $\omega_{\rm geq} = \omega_0 
\exp(-\beta \Omega \psi_{\rm geq})$, the collision term completely
converge to zero all over the system and the Einstein relation is
obtained \cite{Chavanis2001}.
The obtained collision term satisfies the H theorem
which guarantees that the system relaxes to a global equilibrium state.

The organization of this paper is as follows.
In Sec. II, the point vortex system and the Klimontovich formalism are briefly introduced.
In Sec. III, we demonstrate explicit formulae for the diffusion and the drift
terms as intermediate results.
In Sec. IV, a detailed calculation of the diffusion term is shown.
As the similar calculation can be applied to the drift term, 
details for the drift term are omitted.
Finally in Sec. V, three good properties of the collision term are demonstrated.
\section{Point vortex system}
Consider a 2D system consisting of $N$ positive point vortices.
The circulation of each point vortex is given by a positive constant $\Omega$.
\begin{equation}
	\hwz = \sum_i^{N} \Omega \delta(\vr - \vr_i), \label{eqn:point-vortex}
\end{equation}
where $\hwz = \hwz(\vr,t)$ is the $z$-component of the microscopic vorticity on the $x-y$ plane,
and $\delta(\vr)$ is the Dirac delta function in two dimensions.
The microscopic variables in the microscopic equation are
identified by $\hat{\cdot}$.
For brevity, we shall omit the $t$ and $\vr$ dependences, if there is no ambiguity.
Vector $\vr_i = \vr_i(t)$ is the position vector of the $i$-th point vortex.
The discretized vorticity (\ref{eqn:point-vortex}) is a formal solution
of the microscopic Euler equation (\ref{eqn:euler}).
Other microscopic variables are defined by
\begin{eqnarray}
	\hat{\u} & = & \hat{\u}(\vr,t) = -\hat{\vec{z}} \times \grad \hat{\psi}, \\
	\hat{\psi} & = & \hat{\psi}(\vr,t) = \sum_i\Omega_iG(\vr - \vr_i), \\
	G(\vr) & = & -\frac{1}{2\pi} \ln|\vr|,
\end{eqnarray}
where $\hat{\u}$ and $\hat{\psi}$ are the velocity field and the stream 
function in the 2D plane, 
$\hat{\vec{z}}$ is the unit vector in the $z$-direction, 
and $G(\vr)$ is the 2D Green function for the Laplacian
operator with an infinite domain.
As a solution of a macroscopic fluid equation should be given by a smooth function,
the singular solution (\ref{eqn:point-vortex}) should be regarded as
not a solution of the macroscopic equation but one of the microscopic equation.
Thus we call the equation that has the microscopic point vortex solution (\ref{eqn:point-vortex}), 
``microscopic" Euler equation.

Here let us briefly introduce the Klimontovich formalism in plasma physics \cite{Klimontovich}.
It is a concise method to derive the Fokker-Planck equation 
for a macroscopic phase space density $f$
\begin{equation}
	\frac{\partial f}{\partial t} + \vec{v}\cdot \grad_{\vec{r}}f + \frac{q}{m}(\vec{E} + \vec{v} \times \vec{B})\cdot \grad_{\vec{v}}f = \grad_{\vec{v}}\cdot\left({\sf D} \cdot \grad_{\vec{v}} f + \vec{A}f\right),\label{eqn:Fokker-Planck}
\end{equation}
from the Klimontovich equation for a microscopic phase space density $\hat{f}$.
As the dynamics of plasmas is usually dominated by not a collision but a collective behavior due
to long-range interactions, collision term can be neglected and it yields the simplest
form of the kinetic equation called the Vlasov equation:
\begin{equation}
	\frac{\partial f}{\partial t} + \vec{v}\cdot \grad_{\vec{r}}f + \frac{q}{m}(\vec{E} + \vec{v} \times \vec{B})\cdot \grad_{\vec{v}}f = 0.
\end{equation}

We assume that the same hierarchy exists in the 2D fluid equation.
The most microscopic equation is the microscopic Euler equation (\ref{eqn:euler}),
which have the discrete particle solution (\ref{eqn:point-vortex}).
Dividing the microscopic variables into a macroscopic and a fluctuation part, and 
taking ensemble average may yield a macroscopic fluid equation with a collisional
effect like the Fokker-Planck equation.
Ignoring the collision term in the above macroscopic equation, we obtain 
the inviscid fluid equation, namely the macroscopic Euler equation.

Starting equation is the microscopic Euler equation (\ref{eqn:euler}).
Inserting the following expressions into Eq. (\ref{eqn:euler}),
\begin{eqnarray}
	\hwz          & = & \wz     + \delta \wz, \label{eqn:micro-w}\\
	\hat{\vec{u}} & = & \vec{u} + \delta \vec{u},\label{eqn:micro-u}
\end{eqnarray}
and taking the ensemble average, Eq. (\ref{eqn:macroeuler}) is rewritten as the following macroscopic 
equation with the collision term $C=C(\vr, t)$
\begin{equation}
	\frac{\partial }{\partial t} \wz + \div(\u \wz) = C, \label{eqn:collision-term-org}
\end{equation}
\begin{eqnarray}
	C & \equiv & -\div \vec{\Gamma}(\vr, t), \\
	\vec{\Gamma} & = & \langle \delta\u\delta \wz\rangle \nonumber\\
		& = & -\int d\vr' \vec{F}(\vr - \vr') \langle \dwz' \dwz\rangle, \label{eqn:diffusion1}\\
	\vec{F}(\vr) & = & \hat{\vec{z}} \times \grad G(\vr),\\
\end{eqnarray}
where $\vec{\Gamma}$ denotes a diffusion flux.
We note $\dwz'$ for $\dwz(\vr',t)$.
Similarly, we shall note $\wz'$ for $\wz(\vr',t)$.
The following relation is utilized.
\begin{equation}
	\delta \vec{u} =-\int d\vr' \vec{F}(\vr-\vr')\dwz' \label{eqn:delta-u}
\end{equation}
In the next section, we will analytically assess the collision term $C$.
\section{Evaluation of collision term}
We expect that the collision term $C$ appearing in Eq. (\ref{eqn:collision-term-org})
for the point vortex system has two terms, a diffusion term proportional to $\grad \omega$ and 
a drift term proportional to $\omega$,
namely
\begin{equation}
	\vec{\Gamma} \equiv -{\sf D} \cdot \grad \omega + \vec{V} \omega \label{eqn:diffusion_flux}
\end{equation}
where ${\sf D} = {\sf D}(\vr,t)$ is a diffusion tensor, 
and $\vec{V} = \vec{V}(\vr,t)$ is a drift velocity.
To evaluate $\sf D$ and $\vec{V}$ explicitly, we introduce a small parameter $\epsilon$.
We consider a point vortex system with large $N$ keeping the total circulation
$N\Omega$ constant.
The magnitude of $\epsilon$ is of the order of either $1/N$ or $\Omega$.
Orders of the other quantities are:
\begin{eqnarray}
	& & C \approx O(\epsilon^3),\quad {\sf D} \approx O(\epsilon),\quad \vec{V}\approx O(\epsilon^2),\nonumber\\
	& & \frac{\partial \u}{\partial t} \approx O(\epsilon), \quad \frac{\partial \omega}{\partial t} \approx O(\epsilon)\nonumber\\
	& & \omega \approx \grad^2 \psi \approx O(\epsilon^0), \quad \grad \omega \approx O(\epsilon),\nonumber\\
	& & \vec{u} \approx \grad \psi \approx O(\epsilon^0),\quad \grad \vec{u} \approx \grad^2\psi \approx O(\epsilon^0),\nonumber\\
	& & \grad\grad\vec{u}\approx O(\epsilon).
\end{eqnarray}
The expansion parameter $\epsilon$ is the same as one introduced by Chavanis 
in Refs. \cite{Chavanis2001,Chavanis2012} and the references therein.
Expressing $\sf D$ and $\vec{V}$ in the form of a perturbation expansion and gathering
the first order terms for $\sf D$ and the second order terms for $\vec{V}$, an analytical
formula for the collision term $C$ will be obtained.

To rewrite the collision term in Eq. (\ref{eqn:collision-term-org}) according to
the above prospect, 
we introduce a linearized equation obtained by inserting 
Eqs. (\ref{eqn:micro-w}) and (\ref{eqn:micro-u})
into Eq. (\ref{eqn:euler}) and assembling the first-order fluctuation terms:
\begin{equation}
	\frac{\partial }{\partial t} \dwz + \div(\u\dwz) = -\delta\u \cdot \grad \wz. \label{eqn:linearlized}
\end{equation}
As the macroscopic quantities $\vec{u}$ appearing in the second term in the 
left-hand side and $\grad \wz$ in the right-hand side are supposed to be 
constant in the time scale of the microscopic fluctuation,
Eq. (\ref{eqn:linearlized}) can be integrated:
\begin{eqnarray}
	\dwz & = & -\int_{t_0}^t d\tau \delta\u(\vr - \u(t-\tau),\tau) \cdot \grad\wz \nonumber\\
		& & + \dwz(\vr - \u(t - t_0),t_0), \label{eqn:integrated-linear-eq1}\\
	\dwz' & = & -\int_{t_0}^t d\tau \delta\u(\vr' - \u'(t-\tau),\tau) \cdot \grad'\wz' \nonumber\\
		& & + \dwz(\vr' - \u'(t - t_0),t_0), \label{eqn:integrated-linear-eq2}
\end{eqnarray}
where $\grad'\wz' = \grad_{\vr'}\wz(\vr')$.
We call this approximation ``straight-line approximation"
where the trajectory of the point vortex is straight.
The value of $t_0$ is chosen to satisfy $t - t_0 \gg t_c$ where $t_c$ is a 
correlation time of the fluctuation.
Substituting Eqs. (\ref{eqn:integrated-linear-eq1}) and (\ref{eqn:integrated-linear-eq2})
into the correlation term in Eq. (\ref{eqn:diffusion1}), we obtain
\begin{eqnarray}
	& & \left\langle \dwz' \dwz\right\rangle \nonumber\\
	& = & \left\langle \left( -\int_{t_0}^t d\tau \delta\u(\vr' - \u'(t-\tau),\tau) \cdot \grad' \wz' \right) \right.\nonumber\\
	& & 	\times \left.\left( -\int_{t_0}^t d\tau \delta\u(\vr - \u(t-\tau),\tau) \cdot \grad \wz \right) \right\rangle \nonumber\\
	&  & + \left\langle \left( -\int_{t_0}^t d\tau \delta\u(\vr' - \u'(t-\tau),\tau) \cdot \grad' \wz' \right) 
		\dwz(\vr - \u(t-t_0),t_0) \right\rangle \nonumber\\
	&  & + \left\langle \dwz(\vr' - \u'(t-t_0),t_0) 
		\left( -\int_{t_0}^t d\tau \delta\u(\vr - \u(t-\tau),\tau) \cdot \grad \wz \right)\right\rangle \nonumber\\
	&  & + \left\langle \dwz(\vr' - \u'(t-t_0),t_0) \dwz(\vr - \u(t-t_0),t_0) \right\rangle \label{eqn:diffusion2}\\
	& = & -\int_{t_0}^t d\tau \left\langle \delta \u(\vr - \u(t-\tau) ,\tau) \dwz' \right\rangle \cdot \grad \wz \nonumber\\
	& & - \int_{t_0}^t d\tau \left\langle \delta \u(\vr' - \u'(t-\tau) ,\tau) \dwz \right\rangle \cdot \grad' \wz' + O(\epsilon^3) \label{eqn:diffusion3}\\
	& \approx & \int_{t_0}^t d\tau \int d\vr'' \vec{F}(\vr - \u(t-\tau)- \vr'') \cdot \grad \wz \left\langle \dwz(\vr'',\tau) \dwz'\right\rangle  \nonumber\\
	& &  +\int_{t_0}^t d\tau \int d\vr'' \vec{F}(\vr' - \u'(t-\tau)-\vr'') \cdot \grad' \wz' \left\langle \dwz(\vr'',\tau) \dwz\right\rangle. \label{eqn:diffusion4}
\end{eqnarray}
When obtaining formula (\ref{eqn:diffusion3}), we assume that the first 
term in formula (\ref{eqn:diffusion2}) is negligible as it has two nablas.
We drop the last term as it should have a factor of $1/(t-t_0)$ and we focus on $t-t_0\gg t_c$ case.
The time is shifted from $t_0$ to $t$ using the straight-line approximation.
When rewriting formula (\ref{eqn:diffusion3}) as (\ref{eqn:diffusion4}), 
Eq. (\ref{eqn:delta-u}) is used.

Inserting Eq. (\ref{eqn:diffusion4}) into Eq. (\ref{eqn:diffusion1}), the following intermediate results
are obtained:
\begin{eqnarray}
	& & {\sf D} \cdot \grad \wz \nonumber\\
	& = & \int_{t_0}^t d\tau\int d\vr'\int d\vr''\vec{F}(\vr-\vr') \vec{F}(\vr - \u(t-\tau) - \vr'') \cdot \grad \wz \nonumber\\
	& & 	\times \left\langle \dwz(\vr'',\tau) \dwz'\right\rangle, \label{eqn:diffusion}\\
	& & \vec{V}\wz\nonumber\\
	& = & -\int_{t_0}^t d\tau\int d\vr'\int d\vr''\vec{F}(\vr-\vr') \vec{F}(\vr' - \u'(t-\tau) - \vr'') \cdot \grad' \wz' \nonumber\\
	& & 	\times \left\langle \dwz(\vr'',\tau) \dwz\right\rangle. \label{eqn:friction}
\end{eqnarray}
It should be noted that the diffusion term can be expressed in the following
modified Kubo formula:
\begin{equation}
	{\sf D} = \int_{t_0}^t d\tau \langle \delta\u(\vr,t)\delta\u(\vr-\u(t-\tau),\tau) \rangle.
\end{equation}
\section{Evaluation of diffusion and drift terms}
As the expression of the diffusion term (\ref{eqn:diffusion}) is very similar to that of the drift term
(\ref{eqn:friction}), the detailed derivation for diffusion term only is shown:
\begin{eqnarray}
	& & \left\langle \dwz(\vr'',\tau) \dwz'\right\rangle\nonumber\\
	& = & \left\langle \left[ \hwz(\vr'',\tau) -\wz(\vr'',\tau) \right] \left[ \hwz' - \wz' \right] \right\rangle\nonumber\\
	& = & \left\langle \hwz(\vr'',\tau) \hwz'\right \rangle - \wz(\vr'',\tau) \wz' \nonumber\\
	& = & \left\langle \sum_{i=1}^{N} \Omega^2 \delta(\vr''-\vr_i(\tau))\delta(\vr'-\vr_i(t))\right\rangle\nonumber\\
	& & + \left\langle \sum_{i=1}^{N} \sum_{j\neq i}^{N}\Omega^2 \delta(\vr''-\vr_i(\tau)) \delta(\vr'-\vr_j(t))\right\rangle\nonumber\\
	& & - \wz(\vr'',\tau) \wz'.\label{eqn:diffusion6}
\end{eqnarray}
The first term in the last result in Eq. (\ref{eqn:diffusion6}) corresponds
the case of $i=j$,
and the second term corresponds the case of $i\neq j$.

For the $i=j$ case, the formula is rewritten as
\begin{eqnarray}
	& & \left\langle \sum_{i=1}^{N} \Omega^2 \delta(\vr''-\vr_i(\tau))\delta(\vr'-\vr_i(t))\right\rangle \nonumber\\
	& = & \left\langle \sum_{i=1}^{N} \Omega^2 \delta(\vr'' - \vr_i(\tau) - \vr' + \vr_i(t)) \delta(\vr'-\vr_i(t))\right\rangle\nonumber\\
	& = & \sum_{i=1}^{N} \Omega^2 \left\langle \delta(\vr'' - \vr_i(\tau) - \vr' + \vr_i(t)) \delta(\vr'-\vr_i(t))\right\rangle. \label{eqn:diffusion7}
\end{eqnarray}
Here we introduce a stochastic process to evaluate $\vr_i(t) - \vr_i(\tau)$:
\begin{eqnarray}
	\vr_i(t)-\vr_i(\tau) & = & \int_{\tau}^t\u(\vr(\tau'),\tau')d\tau' + \vec{\xi} \nonumber\\
	& \approx & \u'(t-\tau) + \vec{\xi}. \label{eqn:diffusion8}
\end{eqnarray}
The first term in Eq. (\ref{eqn:diffusion8}) represents the straight-line
approximation and the second term represents a Brownian motion.
The stochastic process represented by $\langle \cdot \rangle_{\xi}$ includes all the possible motion
to reach position $\vr_i$ at time $t$.
Then, Eq. (\ref{eqn:diffusion7}) can be rewritten as
\begin{eqnarray}
	& & \sum_{i=1}^{N} \Omega^2 \left\langle \delta(\vr'' - \vr_i(\tau) - \vr' + \vr_i(t)) \delta(\vr'-\vr_i(t))\right\rangle \nonumber\\
	& = & \sum_{i=1}^{N} \Omega^2 \left\langle \delta(\vr''-\vr'+\u'(t-\tau) + \vec{\xi})\right\rangle_{\xi} \left\langle \delta(\vr' - \vr_i(t))\right\rangle\nonumber\\
	& = & \left\langle \delta(\vr''-\vr' + \u'(t-\tau)+ \vec{\xi})\right\rangle_{\xi}\Omega \wz'.
\end{eqnarray}

For the $i\neq j$ case, we introduce an approximation that correlation between the particles 
can be neglected
\begin{eqnarray}
	& & \sum_{i}^{N} \sum_{j \neq i}^{N} \Omega^2 \left\langle \delta(\vr'' - \vr_i(\tau)) \delta(\vr' - \vr_j(t))\right\rangle\nonumber\\
	& \approx & \sum_i^{N} \sum_{j \neq i}^{N} \Omega^2 \left\langle \delta(\vr''-\vr_i(\tau))\right\rangle \left\langle \delta(\vr' - \vr_j(t)) \right\rangle. \label{eqn:diffusion9}
\end{eqnarray}
Also we assume the followings:
\begin{eqnarray}
	\sum_i^{N}\Omega \left\langle \delta(\vr''-\vr_i(\tau))\right\rangle & = & \wz(\vr'',\tau) \label{eqn:approx1}\nonumber\\
	& = & N\Omega\left\langle \delta(\vr''-\vr_i(\tau))\right\rangle.\label{eqn:approx2}
\end{eqnarray}
Inserting Eq. (\ref{eqn:approx2}) 
into Eq. (\ref{eqn:diffusion9}), we obtain
\begin{eqnarray}
	& & \sum_i^{N} \sum_{j \neq i}^{N} \Omega^2 \left\langle \delta(\vr''-\vr_i(\tau))\right\rangle \left\langle \delta(\vr' - \vr_j(t)) \right\rangle \nonumber\\
	& = & \left( \sum_{i=1}^{N} \Omega \left\langle \delta(\vr''-\vr_i(\tau))\right\rangle\right) 
		\times \left( \sum_{j \neq i}^{N}\Omega \left\langle \delta(\vr'-\vr_j(t))\right\rangle\right) \nonumber\\
	& = & N\Omega \frac{\wz(\vr'',\tau)}{N\Omega}(N-1)\Omega\frac{\wz'}{N\Omega}\nonumber\\
	& = & \wz(\vr'',\tau) \wz' - \frac{1}{N}\wz(\vr'',\tau)\wz'.
\end{eqnarray}
Combining the results of $i=j$ and $i\neq j$ cases, we rewrite Eq. (\ref{eqn:diffusion6}) as
\begin{eqnarray}
	& & \left\langle \dwz(\vr'',\tau) \dwz(\vr',t)\right\rangle\nonumber\\
	& = & \Omega \left\langle \delta(\vr'' - \vr' + \u'(t-\tau) + \vec{\xi}) \right\rangle_{\xi} \wz' \nonumber\\
	& & - \frac{1}{N} \wz(\vr'',\tau)\wz'. \label{eqn:diffusion10}
\end{eqnarray}

The two terms in the right hand side of Eq. (\ref{eqn:diffusion10}) are the same order terms as 
we request the total circulation $N\Omega$ constant.
To proceed with the evaluation, a conservation law is introduced
\begin{equation}
	\int d\vr' \left\langle \dwz(\vr'',\tau) \dwz'\right\rangle = 0. \label{eqn:conservation-law}
\end{equation}
Inserting Eq. (\ref{eqn:diffusion10}) into Eq. (\ref{eqn:conservation-law}), we obtain
\begin{eqnarray}
	& & \int d\vr' \left[ \Omega \left\langle \delta(\vr'' - \vr' + \u'(t-\tau) + \vec{\xi}) \right\rangle_{\xi} \wz'\right. \nonumber\\
	& & -\left.\frac{1}{N}\wz(\vr'',\tau) \wz'\right] \nonumber\\
	& = & \Omega \int d\vr' \left\langle \delta(\vr''-\vr'+\u'(t-\tau) + \vec{\xi})\right\rangle_{\xi}\wz' \nonumber\\
	& & -\frac{1}{N}\wz(\vr'',\tau)\int d\vr' \wz' \nonumber\\
	& = & \Omega \int d\vr' \left\langle \delta(\vr''-\vr'+\u'(t-\tau) + \vec{\xi})\right\rangle_{\xi} \wz' \nonumber\\
	& & - \frac{1}{N}\wz(\vr'',\tau)N\Omega \nonumber\\
	& = & 0.
\end{eqnarray}
This equation yields 
\begin{equation}
	\wz(\vr'',\tau) = \int d\vq'\left\langle \delta(\vr''-\vq'+\u(\vq')(t-\tau) + \vec{\xi})\right\rangle_{\xi}\wz(\vq',t) \label{eqn:wr''}\\
\end{equation}
where $d\vr'$ is replaced by $d\vq'$ to avoid ambiguity.
This equation enables that all the quantities at $\tau$ is converted by ones at $t$.
Inserting Eqs. (\ref{eqn:diffusion10}) and (\ref{eqn:wr''}) into Eq. (\ref{eqn:diffusion}),
we obtain
\begin{eqnarray}
 	& & {\sf D} \cdot \grad \wz \nonumber\\
	& = & \Omega \int_{t_0}^t d\tau\int d\vr'\int d\vr''\vec{F}(\vr-\vr') \vec{F}(\vr - \u(t-\tau) - \vr'') \cdot \grad \wz  \nonumber\\
	& & \times \left\langle \delta(\vr''-\vr' + \u'(t-\tau)+ \vec{\xi})\right\rangle_{\xi} \wz'\nonumber\\
	& & - \frac{1}{N} \int_{t_0}^t d\tau \int d\vr' \int d\vr'' \vec{F}(\vr - \vr') \vec{F}(\vr - \u(t-\tau) - \vr'')\cdot \grad \wz\nonumber\\
	& & \times \wz' \int d\vq' \left\langle \delta(\vr''-\vq'+\u(\vq')(t-\tau) + \vec{\xi})\right\rangle_{\xi}\wz(\vq',t). \label{eqn:diffusion11}
\end{eqnarray}
We proceed with the evaluation of the second term in Eq. (\ref{eqn:diffusion11}):
\begin{eqnarray}
	& & - \frac{1}{N} \int_{t_0}^t d\tau \int d\vr' \int d\vr'' \vec{F}(\vr - \vr') \vec{F}(\vr - \u(t-\tau) - \vr'')\cdot \grad \wz\nonumber\\
	& & \times \wz' \int d\vq' \left\langle \delta(\vr''-\vq'+\u(\vq')(t-\tau) + \vec{\xi})\right\rangle_{\xi}\wz(\vq',t) \nonumber\\
	& = & - \frac{1}{N} \int d\vr' \vec{F}(\vr - \vr') \int_{t_0}^t d\tau \int d\vq'\left\langle \vec{F}(\vr - \vq' - (\u - \u(\vq'))(t-\tau) + \vec{\xi})\right\rangle_{\xi} \cdot \grad \wz\nonumber\\
	& & \qquad \times \wz' \wz(\vq',t) \label{eqn:diffusion12}\\
	& = & - \frac{1}{N} \int d\vr' \vec{F}(\vr - \vr') \int_{t_0}^t d\tau \int \frac{d\vk}{(2\pi)^2} \frac{\hat{\vec{z}} \times i\vk}{|\vk|^2} \cdot \grad \wz \nonumber\\
	& & \times\int d\vq' \exp\left[i\vk \cdot (\vr - \vq' - (\u - \u(\vq'))(t-\tau))\right] \left\langle \exp(i\vk\cdot\vec{\xi})\right\rangle_{\xi}\nonumber\\
	& & \quad \times \wz' \wz(\vq', t). \label{eqn:diffusion13}
\end{eqnarray}
To rewrite formula (\ref{eqn:diffusion12}) as (\ref{eqn:diffusion13}), we use the Fourier transformation:
\begin{eqnarray}
	& & \vec{F}(\vr - \vq' - (\u - \u(\vq'))(t-\tau)) \nonumber\\
	& = & \frac{1}{(2\pi)^2} \int d\vk\frac{\hat{\vec{z}}\times i\vk}{|\vk|^2}\exp(i\vk\cdot(\vr - \vq' - (\u - \u(\vq'))(t-\tau))).\label{eqn:fourier}
\end{eqnarray}
The term $\left\langle \exp(i\vk\cdot\vec{\xi})\right\rangle_{\xi}$ represents a Brownian motion
of the point vortices with diffusion tensor ${\sf D}$ and is evaluated by the cumulant expansion:
\begin{eqnarray}
	\left\langle \exp(i\vk\cdot\vec{\xi})\right\rangle_{\xi} 
		& = & \exp\left(-  \frac{\vk \cdot {\sf D} \cdot \vk}{2}(t-\tau)\right) \nonumber\\
	& \equiv & \exp(-\nu(t-\tau))
\end{eqnarray}
where $\nu$ is a small positive parameter.
Inserting the following formula into Eq. (\ref{eqn:diffusion13})
\begin{eqnarray}
	& & \int_{t_0}^t d\tau \exp[-i\vk\cdot(\u - \u(\vq')- \nu)(t-\tau)]\nonumber\\
	& \approx & \pi\delta(\vk \cdot (\u - \u(\vq'))) - \frac{i\vk\cdot (\u - \u(\vq'))}{|\vk\cdot(\u-\u(\vq'))|^2 + \nu^2},
\end{eqnarray}
we obtain
\begin{eqnarray}
	&  & - \frac{1}{N} \int d\vr' \vec{F}(\vr - \vr') \int_{t_0}^t d\tau \int \frac{d\vk}{(2\pi)^2} \frac{\hat{\vec{z}} \times i\vk}{|\vk|^2} \cdot \grad \wz \nonumber\\
	& & \times\int d\vq' \exp\left[i\vk \cdot (\vr - \vq' - (\u - \u(\vq'))(t-\tau))\right] \left\langle \exp(i\vk\cdot\vec{\xi})\right\rangle_{\xi}\nonumber\\
	& & \quad \times \wz'\wz(\vq', t) \nonumber\\
	& = & - \frac{1}{N} \int d\vr' \vec{F}(\vr - \vr') \int \frac{d\vk}{(2\pi)^2} \frac{\hat{\vec{z}} \times i\vk}{|\vk|^2}\cdot \grad \wz \nonumber\\
	&  & \times\int d\vq' \left[ \pi\delta(\vk\cdot(\u-\u(\vq')) - \frac{i\vk\cdot(\u-\u(\vq'))}{|\vk\cdot(\u-\u(\vq'))|^2 + \nu^2}\right]\exp(i\vk\cdot(\vr - \vq'))\nonumber\\
	& & \quad \times \wz' \wz(\vq',t).\label{eqn:diffusion14}
\end{eqnarray}

We substitute $\vr+\vq''$ for $\vq'$ and expand $\vec{u}(\vq')$ and $\wz(\vq')$ 
in the form of Taylor series and retain the zero-th order terms only:
\begin{eqnarray}
	\u(\vq') & = & \u(\vr) + \vq''\cdot \grad \u(\vr) + O(\epsilon), \label{eqn:taylor1}\\
	\wz(\vq') & = & \wz(\vr+\vq'') = \wz(\vr) + O(\epsilon).\label{eqn:taylor2}
\end{eqnarray}
Inserting Eqs. (\ref{eqn:taylor1}) and (\ref{eqn:taylor2}) into Eq. (\ref{eqn:diffusion14}),
we finally obtain
\begin{eqnarray}
	& & - \frac{1}{N} \int d\vr' \vec{F}(\vr - \vr') \int \frac{d\vk}{(2\pi)^2} \frac{\hat{\vec{z}} \times i\vk}{|\vk|^2}\cdot \grad \wz \nonumber\\
	&  & \times \int d\vq' \left[ \pi\delta(\vk\cdot(\u-\u(\vq')) - \frac{i\vk\cdot(\u-\u(\vq'))}{|\vk\cdot(\u-\u(\vq'))|^2 + \nu^2}\right]\exp(i\vk\cdot(\vr - \vq'))\nonumber\\
	& & \quad \times \wz' \wz(\vq',t) \nonumber\\
	& = & - \frac{1}{N} \int d\vr' \vec{F}(\vr - \vr') \int \frac{d\vk}{(2\pi)^2} \frac{\hat{\vec{z}} \times i\vk}{|\vk|^2}\cdot \grad \wz\nonumber\\
	& & \times \int d\vq''\left[ \pi\delta(-\vk \cdot (\vq''\cdot\grad) \u) - \frac{i\vk\cdot(\vq''\cdot \grad)\u}{|\vk \cdot (\vq'' \cdot \grad)\u|^2 + \nu^2}\right]\exp(-i\vk \cdot \vq'')\nonumber\\
	& & \quad \times \wz' \wz. \label{eqn:diffusion15}
\end{eqnarray}
It is found that Eq. (\ref{eqn:diffusion15}) substituting $\vec{k} = -\vec{k}$ and $\vq'' = -\vq''$
changes its sign.
Thus it is concluded that the integral equals zero, i.e. the second term in Eq. (\ref{eqn:diffusion11})
has zero contribution 
and only the first term remains.
The obtained formula for the diffusion term is as follows:
\begin{eqnarray}
	& & -{\sf D}\cdot \grad \wz\nonumber\\
	& = & \Omega\int d\vr'\vec{F}(\vr - \vr')\int \frac{d\vk}{(2\pi)^2}\exp(i\vk\cdot(\vr - \vr')) \frac{\hat{\vec{z}}\times i\vk}{|\vk|^2}\cdot \wz' \grad \wz\nonumber\\
	& & \times\left[\pi\delta(\vk \cdot (\u - \u')) - \frac{i\vk\cdot(\u-\u')}{|\vk \cdot (\u - \u')|^2 + \varepsilon^2}\right].
\end{eqnarray}
The similar calculation can be adapted for the drift term.
For this case, the following conservation law is used:
\begin{equation}
	 \int d\vr \left\langle\dwz(\vr'',\tau) \dwz(\vr,t) \right\rangle=0.
\end{equation}

The whole result including both diffusion and drift terms are given by
\begin{eqnarray}
	\vec{\Gamma}& =& -{\sf D}\cdot \grad \wz + \vec{V}\wz \nonumber\\
	& = & -\Omega \int d\vr'\int \frac{d\vk}{(2\pi)^2} \int \frac{d\vk'}{(2\pi)^2}
		\exp(i(\vk+\vk')\cdot(\vr - \vr'))\nonumber\\
	& & \times \left[\pi\delta(\vk \cdot (\u - \u')) - \frac{i\vk\cdot(\u-\u')}{|\vk \cdot (\u - \u')|^2 + \nu^2}\right]\nonumber\\
	& & \times \frac{\hat{\vec{z}} \times i\vk'}{|\vk'|^2} \frac{\hat{\vec{z}} \times i\vk}{|\vk|^2} \cdot \left( \wz'  \grad \wz - \wz \grad'\wz'\right)\label{eqn:gamma}
\end{eqnarray}
where we have used Eq. (\ref{eqn:fourier}).
\section{Space-averaged collision term}
Equation (\ref{eqn:gamma}) includes the oscillatory term $\exp(i(\vk+\vk')\cdot(\vr - \vr'))$.
To reveal characteristics of the obtained collision term, 
we need to calculate the space average of the collision term
to drop the high-frequency component.
Space average is calculated over the small rectangular area $\Lambda$ with sides both $2L$
located at $\vec{r}$.
The space average of the diffusion flux $\vec{\Gamma}$ given by Eq. (\ref{eqn:gamma})
is defined by
\begin{equation}
	\langle \vec{\Gamma}\rangle_s \equiv \vec{\Gamma}_s(\vr) = \frac{1}{|\Lambda(\vr)|}\int_{\Lambda(\vr)}d\vr'' \vec{\Gamma}(\vr'').
\end{equation}
We assume that the macroscopic variables such as $\u$ and $\wz$ may be
constant inside $\Lambda(\vr)$ and only the term $\exp(i(\vk+\vk')\cdot(\vr - \vr'))$
should be space-averaged:
\begin{eqnarray}
	& & \langle \exp(i(\vk+\vk')\cdot(\vr - \vr')) \rangle_s\nonumber\\
	& = & \frac{1}{(2L)^2}\int_{-L}^{L} dx'' \int_{-L}^{L} dy'' \exp(i(\vk + \vk')\cdot \vr'') \exp(-i(\vk+\vk')\cdot\vr')\nonumber\\
	&\approx& \left(\frac{\pi}{L}\right)^2 \delta(\vk + \vk') \exp(-i(\vk+\vk')\cdot\vr')\nonumber\\
	& = & \left(\frac{\pi}{L}\right)^2 \delta(\vk + \vk')
\end{eqnarray}
where $\vr'' = (x'', y'')$.
Thus, space-averaged diffusion flux is given by
\begin{eqnarray}
	& & \vec{\Gamma}_s(\vr)\nonumber\\
	& = & - \Omega \left(\frac{\pi}{L}\right)^2 \int d\vr' \int \frac{d\vk}{(2\pi)^4}
			\pi\delta(\vk \cdot (\u - \u')) \nonumber\\
	& & \times \frac{\hat{\vec{z}} \times \vk}{|\vk|^2} \frac{\hat{\vec{z}} \times \vk}{|\vk|^2} \cdot \left( \wz'  \grad \wz - \wz \grad'\wz'\right). \label{eqn:space_average}
\end{eqnarray}
In Eq. (\ref{eqn:space_average}), we omit the imaginary part as the collision term consists of only the real part.
Further integration over $\vec{k}$ in Eq. (\ref{eqn:space_average}) can be performed.
The integral concerning $\vec{k}$ is as follows:
\begin{equation}
	\int d\vk \delta(\vk \cdot (\u - \u')) \frac{(\hat{\vec{z}}\times\vk)(\hat{\vec{z}}\times\vk)}{|\vk^4|} \label{eqn:k-integral}.
\end{equation}
Dividing $\vk$ into the parallel and the perpendicular components and inserting them
into Eq. (\ref{eqn:k-integral}), 
\begin{eqnarray}
	\vk & = & k_{\|} \hat{\vec{n}}_{\|} + k_{\perp}\hat{\vec{n}}_{\perp},\nonumber\\
	\hat{\vec{n}}_{\|} & = & \frac{\u - \u'}{|\u - \u'|}, \nonumber\\
	\hat{\vec{n}}_{\perp} & = & \hat{\vec{z}}\times \hat{\vec{n}}_{\|},
\end{eqnarray}
we obtain
\begin{eqnarray}
	& & \int d\vk \delta(\vk \cdot (\u - \u')) \frac{(\hat{\vec{z}}\times\vk)(\hat{\vec{z}}\times\vk)}{|\vk^4|} \nonumber\\
	& = & \int d k_{\|} \int d k_{\perp} \delta(k_{\|} |\u - \u'|) \nonumber\\
	& & \qquad \times \frac{ [\hat{\vec{z}} \times (k_{\|} \hat{\vec{n}}_{\|} + k_{\perp} \hat{\vec{n}}_{\perp}) ] [\hat{\vec{z}} \times (k_{\|} \hat{\vec{n}}_{\|} + k_{\perp} \hat{\vec{n}}_{\perp})] }{|k_{\|}^2+k_{\perp}^2|^2}\nonumber\\
	& = & \int d k_{\perp} \frac{1}{|\u - \u'|} \frac{1}{k_{\perp}^4}(\hat{\vec{z}}\times k_{\perp}\hat{\vec{n}}_{\perp}) (\hat{\vec{z}}\times k_{\perp}\hat{\vec{n}}_{\perp})\nonumber\\
	& = & \int d k_{\perp} \frac{1}{|\u = \u'|} \frac{1}{k_{\perp}^2} \frac{\u - \u'}{|\u - \u'|} \frac{1}{k_{\perp}^2} \frac{\u - \u'}{|\u - \u'|}\nonumber\\
	& = & \frac{(\u - \u') (\u - \u')}{|\u - \u'|^3}2\left[ -\frac{1}{k_{\perp}} \right]_{k_{\rm min}}^{\infty}\nonumber\\
	& = & \frac{(\u - \u') (\u - \u')}{|\u - \u'|^3} \frac{2}{k_{\rm min}}
\end{eqnarray}
where parameter $k_{\rm min}$ is introduced to regularize a singularity and
determined by the largest wave length that does not exceed a system size,
namely $k_{\rm min} = 2\pi/R$ where $R$ is a characteristic system size 
determined by an initial distribution of the vortices.

Finally, we obtain the following formulae for the diffusion and drift:
\begin{eqnarray}
	\vec{\Gamma}_s(\vr) & \equiv & -{\sf D}_s(\vr)\cdot \grad\wz + \vec{V}_s(\vr)\wz, \label{eqn:final-gamma}\\
	{\sf D}_s & = & \frac{\Omega}{(2\pi)^3} \left( \frac{\pi}{L}\right)^2\frac{1}{k_{\rm min}} \int d\vr' \frac{(\u - \u')(\u-\u')\omega'}{|\u - \u'|^3}, \label{eqn:final-D}\\
	\vec{V}_s & = & -\frac{\Omega}{(2\pi)^3} \left( \frac{\pi}{L}\right)^2\frac{1}{k_{\rm min}} \int d\vr' \frac{(\u - \u')(\u-\u')\cdot \grad'\omega'}{|\u - \u'|^3}. \label{eqn:final-V}
\end{eqnarray}
In Eqs. (\ref{eqn:final-D}) and (\ref{eqn:final-V}), two unknown parameters $L$ and $k_{\rm min}$ remain.
We assume that $L = gR$ where $g$ is a size factor ($g \ll 1$).
If we set $g = 1/4\pi$, Eq. (\ref{eqn:final-gamma}) is rewritten as
\begin{eqnarray}
	\vec{\Gamma}_s(\vr) & = & - \frac{\Omega}{R} \int d\vr' \frac{(\u - \u')(\u-\u')}{|\u - \u'|^3}\nonumber\\
	& & \cdot (\wz'\grad\wz - \wz\grad'\wz'). \label{space-averaged-gamma-with-new-coeff}
\end{eqnarray}
\subsection{Collision term in local and global equilibrium states}
At first, let us examine if the collisional effect (\ref{space-averaged-gamma-with-new-coeff}) 
disappears in a local equilibrium state.
We rewrite Eq. (\ref{space-averaged-gamma-with-new-coeff}) into a symbolic form:
\begin{equation}
	\vec{\Gamma}_s(\vr) = - \frac{\Omega}{R} \int d\vr' \vec{\gamma}[\omega, \psi; \omega', \psi'] \label{eqn:functional}
\end{equation}
where $\vec{\gamma}$ is a functional of $\omega$, $\psi$, $\omega'$ and $\psi'$.
Consider a state where temperature is locally uniform in each small region in the system.
We call this state the local equilibrium state
in which the local equilibrium condition is satisfied:
\begin{equation}
	\omega_{\rm leq} = \omega_0\exp(-\beta\Omega\psi_{\rm leq}). \label{eqn:equilibrium}
\end{equation}
Inserting Eq. (\ref{eqn:equilibrium}) into $\vec{\gamma}$ in Eq. (\ref{space-averaged-gamma-with-new-coeff}),
we find that
\begin{eqnarray}
	& & \vec{\gamma}[\wz_{\rm leq},\psi_{\rm leq}; \wz'_{\rm leq}, \psi_{\rm leq}]\nonumber\\
	& = & \frac{(\u_{\rm leq} - \u'_{\rm leq})}{|\u_{\rm leq} - \u'_{\rm leq}|^3}(\u_{\rm leq} - \u'_{\rm leq})\cdot(\wz'_{\rm leq} \grad \wz_{\rm leq} - \wz_{\rm leq} \grad'\wz'_{\rm leq})\nonumber\\
	& = & -\beta \Omega\wz_{\rm leq}\wz'_{\rm leq} \frac{(\u_{\rm leq} - \u'_{\rm leq})}{|\u_{\rm leq} - \u'_{\rm leq}|^3}(\u_{\rm leq} - \u'_{\rm leq})\cdot(\grad\psi_{\rm leq} - \grad'\psi'_{\rm leq})\nonumber\\
	& = & 0 \label{eqn:equilibrium2}
\end{eqnarray}
where $\u_{\rm leq} = -\hat{\vec{z}} \times \grad \psi_{\rm leq}$ is used.
As $\u_{\rm leq} - \u'_{\rm leq}$ is perpendicular to $\grad\psi_{\rm leq} - \grad'\psi'_{\rm leq}$,
$\vec{\gamma}$ is equal to zero and this result indicates a detailed balance
is achieved.
When the system reaches a global thermal equilibrium state \cite{Joyce}
\begin{equation}
	\omega_{\rm eq} = \omega_0\exp(-\beta\Omega\psi_{\rm eq}), \label{eqn:global-equilibrium}
\end{equation}
we obtain
\begin{eqnarray}
	\grad'\wz'_{\rm eq} & = & \wz'_{\rm eq}\frac{\grad'\wz'_{\rm eq}}{\wz'_{\rm eq}}\nonumber\\
	& = & -\beta \Omega \wz'_{\rm eq}(\grad'\psi'_{\rm eq} - \grad \psi_{\rm eq} + \grad \psi_{\rm eq}).
\end{eqnarray}
As $(\u_{\rm eq} - \u'_{\rm eq}) \cdot (\grad'\psi'_{\rm eq} - \grad \psi_{\rm eq}) = 0$,
the drift term in Eq. (\ref{space-averaged-gamma-with-new-coeff}) is rewritten as
\begin{equation}
	\vec{V}_{s,{\rm eq}} = -\beta \Omega {\sf D}_{s,{\rm eq}} \cdot \grad\psi_{\rm eq}
\end{equation}
which is a counterpart of the Einstein relation \cite{Chavanis2001}.
\subsection{Energy-conservative property of collision term}
Time derivative of the total mean field energy $E$ is given by
\begin{eqnarray}
	\frac{d E}{dt} & = & \frac{1}{2} \int d\vr \int d\vr'G(\vr - \vr')
		\left( \frac{\partial \omega'}{\partial t} \omega + \omega'\frac{\partial \omega}{\partial t}\right) \nonumber\\
	& = & \int d\vr\psi \frac{\partial \omega}{\partial t}, \label{eqn:time-derivative-system-energy}
\end{eqnarray}
where 
\begin{eqnarray}
	E & \equiv & \frac{1}{2} \int d\vr \psi \omega\nonumber\\
	& = & \frac{1}{2} \int d\vr \int d\vr' G(\vr-\vr') \omega'\omega.
\end{eqnarray}
Note that the mean field energy $E$ is different from the system energy $\mathscr{H}$
of the point vortex system
\begin{equation}
	\mathscr{H} = -\frac{1}{4\pi}\sum_i \sum_{j \neq i} \Omega_i \Omega_j \ln|\vr_i - \vr_j|.
\end{equation}
Inserting the space-averaged equation of motion
\begin{equation}
	\frac{\partial \omega}{\partial t} + \div(\u\omega) = -\div \vec{\Gamma}_s
\end{equation}
into Eq. (\ref{eqn:time-derivative-system-energy}), we obtain
\begin{eqnarray}
	\frac{d E}{d t} & = & \int d\vr \psi \left( -\div (\u \omega) - \div \vec{\Gamma}_s \right) \nonumber\\
	& = & \int d\vr \grad \psi \cdot \u\omega + \int d\vr \grad \psi \cdot \vec{\Gamma}_s \nonumber\\
	& = & \int d\vr \grad \psi \cdot \vec{\Gamma}_s \nonumber\\
	& = & -\frac{\Omega}{R}\int d\vr \int d\vr'\grad \psi\cdot \frac{(\u - \u')(\u - \u')}{|\u - \u'|^3} \cdot (\wz'\grad\wz - \wz\grad'\wz'). \label{eqn:dEdt1}
\end{eqnarray}
By permuting the dummy variables $\vr$ and $\vr'$ in Eq. (\ref{eqn:dEdt1})
and taking the half-sum of the resulting expressions, we obtain
\begin{eqnarray}
	\frac{d E}{dt} & = & -\frac{\Omega}{2R}\int d\vr \int d\vr'(\grad \psi - \grad'\psi')\cdot \frac{\u - \u'}{|\u - \u'|^3} \nonumber\\
	& & \times (\u - \u')\cdot (\wz'\grad\wz - \wz\grad'\wz')\nonumber\\
	& = & 0.
\end{eqnarray}
We conclude that the obtained collision term conserves the total mean field energy.
\subsection{H theorem}
The entropy function $S$ is defined by using the H function:
\begin{eqnarray}
	S & = & -k_B H,\label{eqn:entropy}\\
	H & = & \int d\vr\frac{\omega}{\Omega}\ln \frac{\omega}{\Omega} + {\rm const.}\nonumber\\
	& = & \frac{1}{\Omega}\int d\vr\omega\ln\omega - N\ln \Omega + {\rm const.}.
\end{eqnarray}
The time derivative of the $H$ function is given by
\begin{eqnarray}
	\frac{dH}{dt} & = & \frac{1}{\Omega} \int d\vr \left( \frac{\partial \omega}{\partial t} (\ln \omega + 1)\right)\nonumber\\
	& = & \frac{1}{\Omega} \int d\vr (-\div (\u \omega) - \div \vec{\Gamma}_s)(\ln \omega + 1)\nonumber\\
	& = & \frac{1}{\Omega} \int d\vr \u\omega \cdot \grad\ln \omega + \frac{1}{\Omega} \int d\vr \vec{\Gamma}_s \cdot \grad \ln \omega\nonumber\\
	& = & -\frac{1}{\Omega} \int d\vr (\div \u) \omega + \frac{1}{\Omega} \int d\vr \vec{\Gamma}_s \cdot \grad \ln \omega\nonumber\\
	& =& \frac{1}{\Omega} \int d\vr \vec{\Gamma}_s \cdot \grad \ln \omega.\label{eqn:dhdt0}
\end{eqnarray}
Inserting Eq. (\ref{space-averaged-gamma-with-new-coeff}) into Eq. (\ref{eqn:dhdt0}),
we obtain
\begin{eqnarray}
	\frac{dH}{dt} & = & -\frac{1}{R} \int d\vr \int d\vr' \frac{\grad\wz}{\wz} \cdot \frac{(\u-\u')(\u-\u')}{|\u-\u'|^3}\nonumber\\
	& & \cdot(\wz'\grad\wz - \wz\grad'\wz') \label{eqn:dHdt1}
\end{eqnarray}
By permuting the dummy variables $\vr$ and $\vr'$ in Eq. (\ref{eqn:dHdt1})
and taking the half-sum of the resulting expressions, we obtain
\begin{eqnarray}
	\frac{dH}{dt} & = & -\frac{1}{2R} \int d\vr \int d\vr' \frac{1}{\wz\wz'}(\wz'\grad\wz - \wz\grad'\wz') \cdot \frac{\u-\u'}{|\u-\u'|^3}\nonumber\\
	&   & \times (\u - \u')\cdot(\wz'\grad\wz - \wz\grad'\wz')  \nonumber\\
	& = & -\frac{1}{2R} \int d\vr \int d\vr' \frac{1}{\wz\wz'}\frac{|(\u-\u')\cdot(\wz'\grad\wz - \wz\grad'\wz')|^2}{|\u-\u'|^3} \nonumber\\
	& \le & 0\label{eqn:dHdt2}
\end{eqnarray}
The integrand of Eq. (\ref{eqn:dHdt2}) is positive or equal to zero,
and $dH/dt$ is negative or equal to zero.
It is concluded that the entropy function $S$ (\ref{eqn:entropy})
is the monotonically increasing function.
\section{Discussion}
We have demonstrated the simple and explicit formula of the 
Fokker-Planck type collision term for nonaxisymmetric point vortex profile
without the collective effect.
It should be noted that it does not include the implicit term 
like Eq. (26) in Ref. \cite{Chavanis2012}.
The collision term exhibits several important properties:
(a) it includes the nonlocal, long-range interaction;
(b) it conserves the mean field energy; 
(c) it satisfiies the H theorem;
(d) its effect vanishes in each local equilibrium region with the same temperature.
When the system reaches a global equilibrium state, the collision term completely
converge to zero all over the system.

There are several outstanding issues remaining.
First, the final formulae (\ref{eqn:final-D}) and (\ref{eqn:final-V}) includes 
unknown parameters $k_{min}$ and $L$.
Second, the integrals in Eqs. (\ref{eqn:final-D}) for ${\sf D}$
and (\ref{eqn:final-V}) for $\vec{V}$ 
contain the divergent integrand, although the combined term $\langle \vec{\Gamma} \rangle_S=-{\sf D}\cdot\grad \omega + \vec{V}\omega$
is regularized.
A more rigorous justification will be needed for fixing the above two issues.
Finnaly, although our result can be adopted for an axisymmetric case, it does not yield
the same results presented by Dubin, O'Neil \cite{DubinONeil1988,Dubin2003} and Chavanis \cite{Chavanis2001,Chavanis2012}.
A reason remains unknown yet.

\begin{acknowledgments}
This work was supported by JSPS KAKENHI Grant Number 24540400.
\end{acknowledgments}


%

\end{document}